\documentclass[a4paper]{article}
\usepackage[latin1]{inputenc}
\sf
\usepackage{amssymb}

\def\C{\mathbb{C}}
\def\H{\mathbb{H}}

\def\om{\omega}
\def\Om{\Omega}

\def\Ga{\Gamma}

\usepackage{amssymb}

\def\C{\mathbb{C}}
\def\H{\mathbb{H}}

\def\La{\Lambda}

\newtheorem{theo}{Theorem}
\newtheorem{defi}{Definition}
\newtheorem{prop}{Proposition}
\newtheorem{cor}{Corollary}
\newtheorem{lem}{Lemma}
\begin{document}
\title{Differential Graded Cohomology and Lie Algebras of Holomorphic
Vector Fields}     
\author{Publication $[9]$, publi\'e dans Commun. Math. Phys. {\bf 208}
(1999) 521--540\\
\\
        Friedrich Wagemann \\
        Institut Girard Desargues -- ESA 5028 du CNRS\\
	Universit\'e Claude Bernard Lyon-I\\
        43, bd du 11 Novembre 1918 \\
	69622 Villeurbanne Cedex FRANCE\\
        tel.: +33.4.72.43.11.90\\
        fax:  +33.4.72.43.00.35\\
        e-mail: wagemann@desargues.univ-lyon1.fr}
\maketitle

Running title: Cohomology of the Lie Algebra of Holomorphic Vector
Fields


secondary: 18F20, 32G15

\begin{abstract}
This article continues work of B. L. Feigin \cite{Fei} and N. Kawazumi
\cite{Kaw} on the Gelfand-Fuks cohomology of the Lie algebra of
holomorphic vector fields on a complex manifold.
As this is not always an interesting Lie algebra (for example, it is 0 for a
compact Riemann surface of genus greater than 1), one looks for other
objects having locally the same cohomology. The answer are a
cosimplicial Lie algebra and a differential graded Lie algebra (well
known in Kodaira-Spencer deformation theory). We calculate the
corresponding cohomologies and the result is very similar to the
result of A. Haefliger \cite{Hae}, R. Bott and G. Segal \cite{BotSeg}
in the case of ${\cal C}^{\infty}$ vector fields. Applications are in conformal
field theory (for Riemann surfaces), deformation theory and foliation theory.
\end{abstract} 
 
\section*{Introduction}

The continuous cohomology of Lie algebras of ${\cal
C}^{\infty}$-vector fields \cite{BotSeg}, \cite{Fuk} has proven to be
a subject of great geometrical interest: One of its most famous
applications is the construction of the Virasoro algebra as the
universal central extension of the Lie algebra of vector fields on the
circle. 

So there is the natural problem of calculating the continuous
cohomology of the Lie algebra of holomorphic vector fields on a
complex manifold, see \cite{Fei} and \cite{Kaw} for Riemann surfaces. 

This work is a solution of this problem for arbitrary complex
manifolds up to the calculation of the cohomology of spaces of
sections of complex bundles on the manifold - this is very close to
the result for $C^{\infty}$-vector fields. We also show the
relation between the cohomology of the holomorphic vector fields and
the differential graded cohomology of some differential graded Lie
algebra. The method is the one of R. Bott and G. Segal
\cite{BotSeg} - used also by N. Kawazumi \cite{Kaw}, and for the relation
with the differential graded cohomology, based on the article of
B. L. Feigin \cite{Fei}. 

One interest is in compact complex manifolds: Here, the Lie
algebra of holomorphic vector fields seems to be too small to be
interesting - for compact Riemann surfaces of genus $g$ it is of
dimension 3 for $g=0$, 1 for $g=1$ and 0 for $g\geq 2$. However,
treating the holomorphic vector fields as a sheaf rather than taking
brutally global sections proves to reveal a richer cohomology theory,
as first remarked by B. L. Feigin \cite{Fei}. 

We study the relation of the sheaf $Hol$ of Lie algebras of
holomorphic vector fields to the sheaf $\mathfrak{g}$ of vector valued
differential forms of type $(0,q)$ where the the values are in the holomorphic
vector fields. It is called the sheaf of Kodaira-Spencer
algebras and it constitues a sheaf of differential graded Lie algebras
which is a fine sheaf resolution of $Hol$. 

We will calculate differential graded (co)-homology for the
Kodaira-Spencer algebra (i.e. the space of global sections of
$\mathfrak{g}$), also with coefficients.
 
Another important idea of this article is the following:

Let $\mathfrak{h}$ a sheaf of differential graded Lie algebras. There
is a sheaf of  differential graded coalgebras $C_{dg,*}(\mathfrak{h})$
with corresponding sheaf of differential graded Lie algebra homology
$H_{*,dg}(\mathfrak{h})$. This is the sheafified Quillen functor, see
\cite{Qui} and \cite{HinSch}. In the same way, there is a sheaf of differential
graded algebras $C^*_{dg}(\mathfrak{h})$ corresponding to the sheaf of
differential graded cohomology $H^*_{dg}(\mathfrak{h})$  of $\mathfrak{h}$.  

Now assume that $\mathfrak{h}$ is not necessarily fine, but that there is a
morphism $\phi$ to a fine sheaf $\mathfrak{g}$ of differential graded Lie
algebras which is a cohomology equivalence
(i.e. $H^*_{dg}(\Ga(U,\mathfrak{g})) = H^*_{dg}(\Ga(U,\mathfrak{h}))$)
on each contractible open set $U$. 

In this case, hypercohomology (for the differential sheaf
$C^*_{dg}(\mathfrak{h})$) and cosimplicial cohomology (i.e. the
cohomology of the realization of the simplicial complex obtained from
applying the functor $C^*_{dg}$ to the Cech resolution of
$\mathfrak{h}$) co\"{\i}ncide under suitable finiteness conditions for
$\mathfrak{g}$ and $\mathfrak{h}$.  

This is true because $\phi$ induces an isomorphism on the cohomology
sheaves of the sheaves $C^*_{dg}(\mathfrak{g})$ and
$C^*_{dg}(\mathfrak{h})$, inducing an isomorphism in
hypercohomology. As $\mathfrak{g}$ is fine, hypercohomology is just
the cohomology of the complex of global sections of
$C^*_{dg}(\mathfrak{g})$. On the simplicial side, we have a morphism
of simplicial cochain complexes induced by $\phi$ which is a
cohomology equivalence on the realizations, see \cite{BotSeg}, lemma
5.9. By a standard argument using partitions of unity for the fine
sheaf $\mathfrak{g}$, see \cite{BotSeg} \S 8, the realization of the
simplicial cochain complex gives the cohomology of the complex of
global sections of $C^*_{dg}(\mathfrak{g})$. 

We will apply this scheme of reasoning to the sheaf of holomorphic
vector fields $\mathfrak{h} = Hol$ on a complex manifold, and its fine
resolution given by the sheaf $\mathfrak{g}$ of $d\bar{z}$-forms with
values in holomorphic vector fields, the sheaf of Kodaira-Spencer 
algebras.

Applications of these calculations are in conformal field theory, cf
\cite{Fei}, in deformation theory, cf \cite{HinSch} and in the theory
of foliations. This work originated in the attempt to
understand Feigin's article \cite{Fei}, so the text is relying
heavily on \cite{Fei}. 

The {\bf content} of the paper reads as follows: The first part is
devoted to cohomology calculations: section 1 is concerned with
the definition of differential graded cohomology (also with
coefficients), hypercohomology, the spectral sequences that go with
them as tools for calculations, and the introduction of the sheaves
$Hol$ and $\mathfrak{g}$; section 2 studies the cohomology of $Hol(U)$
on a Stein open set $U$ linking it with the differential graded cohomology
of $\Ga(U,\mathfrak{g})$; in the end of section 2, we 
treat the cosimplicial version which gives an equivalent point of
view according to the idea explained in the introduction; section 3
gives the calculation of the cohomology in section 2 in terms of the
cohomology of some  spaces of sections of some bundle on the manifold
-  the result is very close to Bott, Haefliger and Segals result
\cite{BotSeg}, \cite{Hae}. The second
part is concerned with the applications of these calculations: section
4 just mentions the existing link to conformal field theory, see
\cite{Fei}; section 5 treats the applications in
deformation theory, following \cite{HinSch}; section 6 shows a glimpse
of possible applications in the theory of characteristic classes of
foliations. 

I am grateful to my thesis advisor C. Roger for encouragement, to the
staff of the mathematical physics section at Lyon University for
discussion, and especially to C. Kapoudjian. Let me also thank
N. Kawazumi and F. Malikov for pointing out some mistakes to me, and
B. L. Feigin for very valuables discussions. 

\section*{Notations}

As a general rule, $g,h$  will denote Lie algebras and gothic letters
$\mathfrak{g},\mathfrak{h}$ will denote sheaves of Lie algebras. For
differential graded Lie algebras, the differential will be displayed
in the notation: $(g,d)$ is a differential graded Lie algebra and
$(\mathfrak{g},d)$ is a sheaf of differential graded Lie algebras.

After the preliminaries, the letter $\mathfrak{g}$ will be reserved
for the Kodaira-Spencer algebra, viewed as a sheaf of
differential graded Lie algebras. 

\section{Preliminaries}

\subsection{Differential Graded (Co-)Homology}

{\bf 1.1.1} Let $g$ an infinite dimensional topological Lie algebra. Its
(co-)homology is calculated by associating to $g$ a differential
graded coalgebra $C_*(g) = (\La^*(g),d)$ and a differential graded
algebra $C^*(g) = (Hom(\La^*(g),\C),d)$, the homological and the
cohomological Chevalley-Eilenberg complex, and then taking their
(co-)homology. In order to keep notations clear, we will suppress
structures we don't need in the notation, as for example the algebra
and coalgebra structure and the grading here. As we deal with tensor
products of infinite dimensional topological vector spaces, we will
always take them to be completed. 

It is worth taking only the continuous duals instead of the algebraic
duals $Hom(\La^*(g),\C)$ in the definition of cohomology, denoted then
$C^*_{cont}(g)$, in order to improve caculability and avoid pathologies. 

{\bf 1.1.2} Let $(g = \bigoplus_{i=0}^{n}g^i, \bar{\partial})$ a
(cohomological) differential graded Lie algebra, dgla for short. There
are as before two functors, noted here $C_{*,dg}$ and $C^*_{dg}$,
associating to $(g,\bar{\partial})$ a differential graded coalgebra
$C_{*,dg}(g)$ and a differential graded algebra $C^*_{dg}(g)$. We will
assume continuous duals in $C^*_{dg}(g)$ without displaying it in the
notation.   

$C_{*,dg}(g)$ and $C^*_{dg}(g)$ extend the functors in 1.1.1: for a
trivial dgla $(g,\bar{\partial}) = (g,0)$ with only its $0^{\rm th}$ space in the
grading non-zero, we have $C_{*,dg}(g) = C_*(g)$ and $C^*_{dg}(g) =
C^*_{cont}(g)$. 

$C_{*,dg}$ is called the Quillen functor, see \cite{Qui}, and
explicitly constructed in \cite{HinSch} \S 2.2. The cohomology version
was, to the knowledge of the author, first used by Haefliger
\cite{Hae}, see also \cite{SchSta} for useful remarks.

{\bf 1.1.3} Explicitly

\begin{displaymath}
C_{k,dg}(g) := \bigoplus_{k=p+q}C_{dg}^p(g)^q :=
\bigoplus_{k=q+p}S^{-p}(g[1])^q  
\end{displaymath}

 as graded vector spaces. ``$dg$'' stands for ``differential
graded''. Here $S^p(g[1])^q$ is the graded symmetric algebra
$S^*$ on the shifted by 1 graded vector space $g[1]$, i.e.

\begin{displaymath}
g[1]^q := g^{q+1}.
\end{displaymath}

Note that for $g^0\not= 0$, we have in $g[1]$ a component of degree
$-1$. $S^{-p}(g[1])^q$ is bigraded by
the tensor degree $-p$ and the internal degree $q$ which is induced
by the grading of $g[1]$. The differential on $C_{*,dg}(g)$ is the
direct sum of the graded homological Chevalley-Eilenberg differential
in the tensor direction (with degree reversed in order to have a
cohomological differential) and the differential induced on
$S^*(g[1])^*$ by $\bar{\partial}$, still noted $\bar{\partial}$.

Note that the differential graded {\it homology} of $g$, denoted by
$C_{*,dg}(g)$, is calculated by a {\it cohomological} complex, but
involving the {\it homological} Chevalley-Eilenberg differential. 

{\bf 1.1.4} $C_{*,dg}(g)$ is the total direct sum complex associated
to the bicomplex $\{S^{-p}(g[1])^q\}_{p,q}$. So, there is  a spectral
sequence associated to the filtration by the columns, taking first
cohomology in one column, i.e. cohomology with respect to
$\bar{\partial}$. Note that $H^*_{\bar{\partial}}$ is a functor from
dgla's to graded Lie algebras. Let us identify the $E_2$ term as well
as where to the sequence converges:

\begin{lem}
Suppose that the complex $(g,\bar{\partial})$ is a
topological complex of Fr\'echet nuclear spaces. 

There is a spectral sequence with 

\begin{displaymath}
E_2^{p,q} = H^p_{gla}(H^q_{\bar{\partial}}(g))
\end{displaymath}

converging to $H_{p+q,dg}(g)$, i.e. the differential graded homology
of $(g,\bar{\partial})$. Here, $H^*_{gla}$ denotes the cohomology of graded
Lie algebras.  
\end{lem}

{\it Remark:} Other names for the morphisms involved in a topological
cochain complex are strict morphisms or homomorphisms, see \cite{Bou}
Ch. III, \S 2, no. 8. These are not necessarily ``morphismes forts''
or split in the sense of \cite{Gui} or \cite{Tay}.

{\it Proof:}

For $E_2$, the only thing which is not clear is 

\begin{displaymath}
C_{*,dg}(H^*_{\bar{\partial}}(g)) = H^*_{\bar{\partial}}(C_{*,dg}(g)).
\end{displaymath}

This follows directly from prop. 2.1 in \cite{Qui} in case we would
not have been taking completed tensor products.

This proposition holds also in the completed tensor product version,
when the spaces involved are Fr\'echet nuclear spaces or its strong
duals, and the complexes are topological:

In this case, there is a K\"unneth formula, cf \cite{Kaw}
p. 673. Then we can conclude as in \cite{Qui}, but we won't have a
topological isomorphism, which is irrelevant for us.

The convergence is a more difficult problem because the shifting
$g[1]$ in $C_{*,dg}(g)$ creates an internal degree -1 for a dgla
$(g=\bigoplus_{i=0}^ng^i,\bar{\partial})$ and so the spectral sequence is
not contained in the third quadrant. Actually, it is contained in the
fourth quadrant. By the classical convergence theorem (cf \cite{Wei}
p. 135) the spectral sequence associated to the filtration by the
columns converges to the total direct sum
complex. This is by definition our differential graded homology.$\square$

{\bf 1.1.5} It is clear how to incorporate coefficients in a
differential graded module $(M =
\bigoplus_{i=0}^kM^i,\tilde{\partial})$: such a module $M$ is given as
the direct sum of its components $M^i$ and carries a differential
$\tilde{\partial}$ and an action of a dgla $(g,\bar{\partial})$ such
that for $x\in g$ and $m\in M$, we have

\begin{displaymath}
\tilde{\partial}(x.m) = \bar{\partial}(x).m +
(-1)^{deg(x)}x.\tilde{\partial}(m).
\end{displaymath}

Now, take the graded tensor product $C_{*,dg}(g)\otimes M$ or the
graded Hom-functor $Hom(C_{*,dg}(g),M)$ with the action incorporated
in the Chevalley-Eilenberg differential and the differential
$\tilde{\partial}$ glued together with the differential $\bar{\partial}$ on
$C_{*,dg}(g)$, i.e. in the homological case

\begin{displaymath}
\partial_{tot}(x\otimes m) = (-1)^{deg
x}(1\otimes\tilde{\partial})(x\otimes m) +
(\bar{\partial}\otimes 1)(x\otimes m)
\end{displaymath}

and in the cohomological case

\begin{displaymath}
\partial_{tot}f = \tilde{\partial}\circ f +
(-1)^{deg f} f\circ \bar{\partial}.
\end{displaymath}

We suppose further that $M$ is a topological
Fr\'echet nuclear module, and take completed tensor products.

Note that, as before, the functor $H^*_{\partial_{tot}}$ transforms
differential graded objects in graded objects. 

There is analoguously a spectral sequence and its corresponding lemma
in this case:

\begin{lem}
There is a spectral sequence with 

\begin{displaymath}
E_2^{p,q} = H^p_{gla}(H^q_{\partial_{tot}}(g\otimes M))
\end{displaymath}

converging to $H_{p+q,dg}(g,M)$, i.e. the differential graded homology
of $(g,\bar{\partial})$ with coefficients in the differential graded module $(M,\tilde{\partial})$. Here, $H^*_{gla}$ denotes the cohomology of graded
Lie algebras with coefficients.  
\end{lem}

{\bf 1.1.6} Now we want to calculate differential graded cohomology
instead of homology, so let me specify a setting where this is
possible.

Let $(g,\bar{\partial})$ a topological dgla such that $g$ is a
Fr\'echet nuclear space. This permits to calculate cohomology by
calculating homology on the continuous dual: 

\begin{lem}
We have

\begin{displaymath}
C^*_{dg}(g) \cong (C_{*,dg}(g^*))
\end{displaymath} 

where we treat all objects as graded vector spaces and $g^*$ is the
continuous dual of $g$ as a topological vector space.
\end{lem}

{\it Proof:}

This follows directly from the following proposition, see for example
\cite{Tre}. prop. 50.7 p. 524:

\begin{prop}
The continuous dual of a completed tensor product of two nuclear
Fr\'echet spaces is the completed tensor product of the continuous
duals of the two spaces.$\square$
\end{prop}

So, there is a spectral sequence for the differential graded
cohomology in case $(g,\bar{\partial})$ is also a topological complex,
namely, the one from lemma 1. 
 
{\bf 1.1.7} In the same setting as in 1.1.6, suppose that
$(g,\bar{\partial})$ is a resolution of a Lie algebra $h$ which is a
topological complex.  As J.- P. Serre showed in \cite{Ser}, the
$\bar{\partial}$ resolutions on compact K\"ahler or Stein manifolds
are always topological cochain complexes. 

For topological cochain complexes of Fr\'echet nuclear (or its dual)
spaces, it is known that the strong dual complex is topological and
has the dual cohomology spaces, cf \cite{Kaw} p. 673. This suites very
well with our approach of cohomology by the homology on the duals.

In conclusion, the resolution $(g,\bar{\partial})$ of $h$ induces an
exact sequence for the strong duals, and by the remark in 1.1.4, the
spectral sequence in cohomology collapses at the second term. 

So, we have:

\begin{lem}  \label{*}
Let $(g,\bar{\partial})$ a dgla as in 1.1.5 such that

\begin{displaymath}
H^*_{\bar{\partial}}(g) = \left\{ \begin{array}{r@{\quad = \quad}l}
                                          h\,\,\,{\mathrm for}\,\,\,* & 0 \\
                                          0\,\,\,{\mathrm for}\,\,\,* &
                                          1,2,\ldots
                             \end{array}   \right.
\end{displaymath}

Then

\begin{displaymath}
H^*_{dg}(g) = H_{cont}^*(h)
\end{displaymath}
\end{lem}

Let me remark that the spectral sequence here is converging in the
sense of complete convergence, cf \cite{Wei} p. 139, and to the total
direct product complex.

{\bf 1.1.8} All these notions extend to sheaves of
Lie\-al\-ge\-bras and sheaves of dgla's: Let $X$ a complex manifold of
complex dimension $n$. Denote by ${\cal O}_X$ the coherent sheaf of
holomorphic functions on $X$ and by ${\cal E}_X$ the sheaf of
$C^{\infty}$ functions on $X$. Let $\mathfrak{g}$ a sheaf of ${\cal
O}_X$-modules which are Lie algebras. Note that the bracket is not a
morphism of ${\cal O}_X$-modules. In some contexts, the
action of the elements of the Lie algebra on $f\in{\cal O}_X$ should
be specified: this leads to the concept of twisted Lie algebras. This
is for example the case when considering tensor products over ${\cal
O}_X$. In our context, everything is $\C$-linear, so we need not
specify this action. In the same way, let
$(\mathfrak{g},\bar{\partial})$ a sheaf of dgla's which are ${\cal
E}_X$-modules. 

We denote by $\Ga(\mathfrak{g})$, $\Ga(X,\mathfrak{g})$ or
$\mathfrak{g}(X)$ the dgla of global sections of the sheaf
$\mathfrak{g}$.

By the previous sections, we can associate to $\mathfrak{g}$ resp. to
$(\mathfrak{g},\bar{\partial})$ sheaves of differential graded coalgebras
$C_*(\mathfrak{g})$, $C_{*,dg}(\mathfrak{g})$, $H_*(\mathfrak{g})$ and
$H_{*,dg}(\mathfrak{g})$ where the last two carry the trivial
differential. In the same way, we have sheaves of differential graded
algebras $C^*_{cont}(\mathfrak{g})$, $C_{dg}^*(\mathfrak{g})$,
$H^*_{cont}(\mathfrak{g})$ and $H_{dg}^*(\mathfrak{g})$.

Furthermore, we have differential graded coalgebras $C_*(\Ga(\mathfrak{g}))$, $C_{*,dg}(\Ga(\mathfrak{g}))$, $H_*(\Ga(\mathfrak{g}))$ and
$H_{*,dg}(\Ga(\mathfrak{g}))$, and the corresponding algebras.

\subsection{Examples}

{\bf 1.2.1} The prescription $U\mapsto Hol(U)$ where $U$ is an open
set of $X$ and $Hol(U)$ is the Lie algebra of holomorphic vector fields
on $U$ is a sheaf of Lie algebras, denoted by $Hol$. It is a coherent
sheaf. It is in some respect the opposite of a fine sheaf: its
restriction maps are injective.

We have a sheaf of differential graded algebras $C^*_{cont}(Hol)$
associated to $Hol$. To be explicit, it is the sheaf ${\cal H}om
(\La^*(Hol),\C)$ of morphisms of sheaves between $\La^*(Hol)$ and the
constant sheaf $\C$. Its underlying presheaf is

\begin{displaymath}
U\mapsto {\cal H}om_{cont} (\La^*(Hol)|_U,\C|_U)
\end{displaymath}

Here, ${\cal H}om_{cont}({\cal F},{\cal G})$ is the functor of continuous sheaf
morphisms between two sheaves of topological spaces ${\cal F}$ et
${\cal G}$ , i.e. of morphisms of presheaves $\phi_U = \{\phi_V : {\cal
F}(V)\to{\cal G}(V)\}_{V\subset U}$ such that every $\phi_V$ is continuous. 

In particular, it is a differential sheaf, and
one subject of this article will be to calculate its hypercohomology:

Taking a sheaf resolution of every graded component of
$C^*_{cont}(Hol)$, we get in a standard way (cf \cite{God} 4.5,
p. 176) a resolution of $C^*_{cont}(Hol)$. This gives a bicomplex; the
cohomology of the total complex associated to it is by definition the
hypercohomology of $C^*_{cont}(Hol)$, denoted by
$\H(X,C^*_{cont}(Hol))$.

As  $C^*_{cont}(Hol)$ is bounded below, we have two converging
spectral sequences (associated to the two canonical filtrations for
the bicomplex) for hypercohomology. We need in 2.1.4 the first one,
the one given by the filtration by the columns. Its $E_2$ term is

\begin{displaymath}
E^{p,q}_2 = H^p(X,H^q_{cont}(Hol)).
\end{displaymath}
  
Here, $H^p(X,{\cal F})$ is the sheaf cohomology of the sheaf ${\cal
F}$.

The second one is given by the filtration by the rows. Its $E_2$ term is

\begin{displaymath}
E^{p,q}_2 = H^q_d(H^p(X,C^*_{cont}(Hol))).
\end{displaymath}
  
Here, $d$ is the differential which $C^*_{cont}(Hol)$ induces in the
resolutions of every component.   

{\bf 1.2.2} Let $E$ a holomorphic vector bundle over the complex
manifold $X$. Denote by ${\cal O}(E)$ the sheaf of (germs of)
holomorphic sections of $E$. Denote by $\Om^{k,l}$ the sheaf of (germs
of $C^{\infty}$ sections of) differential forms of type $(k,l)$ on
$X$. The tensor product $\Om^{0,*}\otimes{\cal O}(E)$ is a sheaf on
$X$, the sheaf of (germs of $C^{\infty}$ sections of) differential
forms with values in $E$ (of type $(0,*)$). Let me denote by
$\mathfrak{g}$ this sheaf for $E = TX$, the complex tangent bundle of
$X$. Note that ${\cal O}(TX)$ is simply $Hol$. 

$\mathfrak{g}$ is a sheaf of dgla's: It is a vector space, graded by
the degree of the differential form. The bracket on every open set is
the restriction to the $(0,*)$-type forms with values in $TX$ of the
Fr\"olicher-Nijenhuis bracket on $\Ga(\Om^{*,*}\otimes Vect)$ where
$Vect$ is the sheaf of all vector fields on the real manifold
underlying $X$. This bracket is explained for example in
\cite{KoMiSl}, see also section 2.2 of the present article. To give a
short indication, it is the bracket of endomorphisms by viewing vector
valued differential forms as derivations of the graded algebra of
differential forms. The differential is just $\bar{\partial}$. It is
easy to see that $\bar{\partial}$ acts as a graded derivation on the
Fr\"olicher-Nijenhuis bracket. We denote by
$(\mathfrak{g},\bar{\partial})$ this sheaf of dgla's.

$\mathfrak{g}$ is a fine sheaf because it is a sheaf of $C^{\infty}$
sections. So its sheaf of dg algebras ${\cal
H}om_{cont}(C_{*,dg}(\mathfrak{g}),\C)$ which is as before a sheaf of
morphisms of sheaves, is in fact isomorphic to the sheaf of morphisms
between the spaces of sections: given a morphism of sheaves
$\phi_U:C_{*,dg}(\mathfrak{g})|_U\to \C|_U$. i.e. a compatible family 

\begin{displaymath}
\phi_U = \{\phi_V: C_{*,dg}(\mathfrak{g})(V)\to \C(V)\}_{V\subset U},
\end{displaymath}

we can construct a morphism of the spaces of global sections on $U$ by
partitions of unity. 

It is well known that the hypercohomology of a fine
differential sheaf is just the cohomology of its complex of global
sections, see for example \cite{God} thm. 4.6.1 p. 178. This implies

\begin{displaymath}
\H(X,C^*_{dg}(\mathfrak{g})) = H^*_{dg}(\Ga(\mathfrak{g})).
\end{displaymath}

Another goal of this article is to calculate the differential graded
cohomology of the Kodaira-Spencer algebra $\Ga(\mathfrak{g})$.

{\bf 1.2.3} There is one remark in order: Actually, we should indicate
in the notation $H^*_{dg}$ the way in which proceeded to take total
complexes associated to double complexes and cohomology with respect
to differentials. The ambiguity involved stems from the fact that we
are considering here hypercohomology of a bicomplex of sheaves, so the
underlying homological problem is a TRI-complex. For example, it would
not be the same to apply the global section functor term by term to
the bicomplex (for example on a compact Riemann surface), and take its
differential graded cohomology afterwards.

Let us thus denote by $^1H^*_{dg}$ the differential graded cohomology
obtained by the hypercohomology of the complex of sheaves given by the
total complex of the bicomplex of sheaves.

Let us also denote by $^2H^*_{dg}$ the differential graded cohomology
obtained by the cohomology of the total complex of the bicomplex of
global sections of the double complex of sheaves. 

{\bf 1.2.4} Let me remark that the sheaf $Hol$ is a sheaf of
topological Fr\'echet nuclear Lie algebras because of the canonical
Fr\'echet topology on the space of sections on a coherent sheaf, see
for example \cite{GraRem} Ch. V, \S 6.

In the same way, $\mathfrak{g}$ is a sheaf of topological Fr\'echet
nuclear dgla's, see for example \cite{Ham}.

$\mathfrak{g}$ as a space of $C^{\infty}$ functions carries the
$C^{\infty}$ topology, and the canonical topology on a space of
holomorphic functions is the same as the one induced from the
$C^{\infty}$ topology on the $C^{\infty}$ functions. 

\section{The cohomological link between $Hol$ and $\mathfrak{g}$}

There is a strong relationship between $Hol$ and $\mathfrak{g}$ based
on the fact that $\mathfrak{g}$ is a fine sheaf resolution of $Hol$. We
will first show this for trivial coefficients, and then construct the
right category of modules such that the relationship still holds for
cohomology with coefficients.

\subsection{Trivial coefficients}

{\bf 2.1.1} Recall some $\bar{\partial}$ resolutions:

\begin{lem}
There is an exact sequence of sheaves
 
\begin{displaymath}
0\to Hol\to (\Om^{0,0}\otimes Hol)\stackrel{\bar{\partial}}{\to} (\Om^{0,1}\otimes Hol) \to \ldots \to
(\Om^{0,n}\otimes Hol) \to 0.
\end{displaymath}
\end{lem}

{\it Proof:}

Actually, we have an exact sequence of sheaves 

\begin{eqnarray*}
0\to\Omega_{hol}^p\otimes_{\cal O}{\cal O}(E)\to\Om^{p,0}\otimes_{\cal O}{\cal O}(E)\stackrel{\bar{\partial}\otimes1}{\to}\Om^{p,1}\otimes_{\cal O}{\cal O}(E)\stackrel{\bar{\partial}\otimes1}{\to}\ldots \\ \ldots\stackrel{\bar{\partial}\otimes1}{\to}\Om^{p,n}\otimes_{\cal O}{\cal O}(E)\to 0
\end{eqnarray*}

with the sheaf of holomorphic differential forms on $X$,
$\Omega_{hol}^p$, for every holomorphic fiber bundle $E$ on $X$, see
for example \cite{Wel}. Taking $p=0$ and $E = TX$, we have our
sequence.$\square$ 

\begin{cor}
The sheaf $(\mathfrak{g},\bar{\partial})$ is a resolution of $Hol$ by fine
sheaves.  
\end{cor}

{\bf 2.1.2} Let us look for the open sets $U$ where the corollary
holds not only for the sheaves, but for the spaces of global sections
on $U$.

\begin{defi}
An open set $U\subset X$ of a complex manifold $X$ is called a Stein
open set, if we have the following vanishing condition on coherent
sheaf cohomology:

\begin{displaymath}
H^*(U,{\cal F}) = 0\,\,\,\,\,\forall\,\,* = 1,2,3,\ldots
\end{displaymath}

and for all coherent sheaves ${\cal F}$ on $X$.
\end{defi} 

\begin{lem}
For every Stein open set $U\subset X$,
$(\Ga(U,\mathfrak{g}),\bar{\partial})$ is a resolution of $Hol(U)$.
\end{lem}

{\it Proof:}

This follows from standard sheaf cohomology theory: As
$(\mathfrak{g},\bar{\partial})$ is a fine sheaf resolution of $Hol$,
$H^*(U,Hol)$ is the cohomology of $(\Ga(U,\mathfrak{g}),\bar{\partial})$. By
definition of $U$, this cohomolgy is 0 except perhaps in degree 0. In
degree 0, it is $Hol(U)$.$\square$


Apply now lemma \ref{*} to get immediatly

\begin{cor}
For every Stein open set $U$, we have an isomorphism

\begin{displaymath}
H^*_{cont}(Hol(U)) \,\,\,=\,\,\, ^1H^*_{dg}(\Ga(U,\mathfrak{g})).
\end{displaymath}
\end{cor}

{\bf 2.1.3} We can state this result in a completely formal setting:

Recall that $W_1$ is the Lie algebra of formal vector fields in 1
variable. Consider  

\begin{displaymath}
G := W_1[[\bar{z},t]]\,/\,(t^2)
\end{displaymath}

Then, $G$ is a differential graded Lie algebra with the bracket:

\begin{displaymath}
[X\bar{z}^kt^n,Y\bar{z}^lt^m] = [X,Y]\bar{z}^{k+l}t^{n+m}
\end{displaymath}

where $X,Y\in W_1$ - it is the usual bracket on a tensor product of a
Lie algebra with an associative algebra. The grading is given by the
polynomial degree in $t$. The differential is just the operator
$\bar{\partial}$ defined by 

\begin{displaymath}
\bar{\partial}(\sum_if_i(z,\bar{z})t^i\frac{\partial}{\partial z}) =
\sum_i\frac{\partial}{\partial \bar{z}}f_i(z,\bar{z})t^{i+1}\frac{\partial}{\partial z}
\end{displaymath}

In particular, the elements of $G$ without $\bar{z}$ are the kernel of
$\bar{\partial}$ - these are the formal holomorphic vector fields. 

So the theorem can be stated in the 1 dimensional formal case as

\begin{theo}
\begin{displaymath}
^1H^*_{dg}(G) \cong H^*_{cont}(W_1) \cong H^*_{sing}(S^3)
\end{displaymath}
\end{theo}

Of course, there exists also the $n$-dimensional version, but it is
too cumbersome to write it down. 

{\bf 2.1.4} We can persue 2.1.2 a little bit further applying
hypercohomology:

\begin{theo}
For every complex manifold $X$, there is an isomorphism

\begin{displaymath}
\H^*(X,C^*_{cont}(Hol)) \,\,\,=\,\,\, ^1H^*_{dg}(\Ga(X,\mathfrak{g})).
\end{displaymath}
\end{theo}

{\it Proof:}

The preceding corollary gives the isomorphism on the filtrant family of
Stein neighbourhoods of a point $x\in X$. Passing to the inductive
limit, we get an isomorphism of the cohomology sheaves

\begin{displaymath}
H^*_{cont}(Hol) \,\,\,=\,\,\,^1H^*_{dg}(\mathfrak{g}).
\end{displaymath}

Recall now the hypercohomolgy spectral sequence from 1.2.1. 

The inclusion sheaf morphism $Hol\to \mathfrak{g}$ gives a morphism of
differential sheaves inducing a morphism of spectral sequences. This
morphism is an isomorphism on the terms $E_2$, so by the standard
comparison theorem for spectral sequences, we have an isomorphism of
the limit terms. 

It remains to recall the result of 1.2.2 stating 

\begin{displaymath}
\H^*(X,C^*_{dg}(\mathfrak{g})) \,\,\,=\,\,\, ^1H^*_{dg}(\Ga(X,\mathfrak{g})).\,\,\square
\end{displaymath}
 
{\bf 2.1.5} Let us remark that there is an analogous situation for the
Hochschild cohomology of the algebra of holomorphic functions ${\cal
O}_X(X)$: we have a fine $\bar{\partial}$-resolution of the sheaf ${\cal
O}_X$ by the sheaves of differential forms of type $(0,k)$,
$\Om^{0,k}$. On a Stein open set $U$, we have an isomorphism between
the Hochschild cohomology of ${\cal O}_X(U)$ and the differential
graded Hochschild cohomology of the differential graded algebra
$(\oplus_{k=0}^n\Om^{0,k}(U),\wedge,\bar{\partial})$. As before, we
can pass to the cohomology sheaves and then to hypercohomology. We can
even have the cosimplicial cohomology - see section 2.3. 

\subsection{The coefficient case}

Note that $Hol(U)$ is a $Hol(U)$-module and
$(\Ga(U,\mathfrak{g}),\bar{\partial})$ is a differential graded
$\Ga(U,\mathfrak{g})$-module by the adjoint action for an open set
$U$. In particular, $(\Ga(U,\mathfrak{g}),$ $\bar{\partial})$ as a
differential graded module $(M,\tilde{\partial})$ verifies 

\begin{displaymath}
\tilde{\partial}(x.m) = \bar{\partial}(x).m +
(-1)^{deg(x)}x.\tilde{\partial}(m),
\end{displaymath}

just by the fact that $\tilde{\partial} = \bar{\partial}$ acts as a graded
derivation on the Fr\"olicher-Nijenhuis bracket. We can write the
Fr\"olicher-Nijenhuis bracket in our case locally as

\begin{eqnarray*}
[\phi\otimes X,\psi\otimes Y]&=&\phi\wedge\psi\otimes[X,Y]\,+\, \\
&-&(i_Y\partial\phi\wedge\psi\otimes X\,-\,(-1)^{kl}i_X\partial\psi\wedge\phi\otimes
Y) \\ 
&=&\phi\wedge\psi\otimes[X,Y]\,+\, \\
&+&\phi\wedge{\cal L}_X\psi\otimes Y\,-\,{\cal L}_Y\phi\wedge\psi\otimes X\,
\end{eqnarray*}

for $\phi\in\Om^{0,k}(U)$, $\psi\in\Om^{0,l}(U)$ and $X,Y\in Hol(U)$.

We will look for
differential graded $\Ga(U,\mathfrak{g})$-modules giving a theorem as
in 2.1 but with coefficients. 

{\bf 2.2.1} Let $E$ be a holomorphic vector bundle on $X$. As before,
let ${\cal O}(E)$ be the sheaf of holomorphic setions of $E$. It has a
resolution by fine sheaves, given explicitly in the proof in section
2.1.1. Denote by ${\cal E}(E)^{0,*}$ the direct sum over all sheaves
in this resolution.

{\bf 2.2.2} As before, $\Ga(U,{\cal E}(E)^{0,*})$ is a differential
graded vector space $(M,\tilde{\partial})$ which is resolution of the
vector space $\Ga(U,{\cal O}(E))$ for any Stein open set $U$. 

{\bf 2.2.3} Let us now suppose that the Lie algebra $Hol(U)$ acts on
$\Ga(U,{\cal O}(E))$ by differential operators - we can speak of a
local action. We can define a differential graded action locally by
setting

\begin{displaymath}
(\phi\otimes X).(\psi\otimes v)\,=\,\phi\wedge\psi\otimes X.v\,+\,\phi\wedge{\cal L}_X\psi\otimes v
\end{displaymath}

where $v\in {\cal O}(E)$. Note that we dropped the term which is not
realizable without an action of $v$ on the forms and an inclusion of
the holomorphic vector fields into ${\cal O}(E)$. It is obvious that
it is in fact a global action. 

So we have constructed a differential graded
$\Ga(U,\mathfrak{g})$-module naturally induced by the action of
$Hol(U)$ on $\Ga(U,{\cal O}(E))$. It is easy to extend this
correspondence to maps between modules, so we have constructed a
category of differential graded modules corresponding to the category
of local $Hol(U)$-modules.

{\bf 2.2.4} We have a functor from local differential graded
modules to the category of local $Hol(U)$-modules simply by taking the
cohomology with respect to $\tilde{\partial}$. So we get an
equivalence of categories between the category of local
$Hol(U)$-modules and a subcategory of the category of differential graded
modules. 

{\bf 2.2.5} Call now the induced module of a local $Hol(U)$-module
either the module constructed in 2.2.3 or - if the $Hol(U)$-module is
$Hol(U)$ itself with the adjoint action - take $\Ga(U,\mathfrak{g})$
with its adjoint action. Unfortunately, we have to make this
distinction because of the difference in the formulae for the action
in 2.2.3 and the adjoint action. 

{\bf 2.2.6} We can now formulate
the analogous theorem in the coefficient case:
 
\begin{theo}
On a Stein open set $U$, a local $Hol(U)$-module $N(U)$ induces a
differential graded module $(M(U),\tilde{\partial})$ which is its
resolution. So we have:

\begin{displaymath}
^1H^*_{dg}(\mathfrak{g}(U),(M(U),\tilde{\partial})) \cong H^*_{cont}(Hol(U),N(U))
\end{displaymath}
\end{theo}

{\it Proof:}

Following 2.2.3, the first statement is clear.

 By the spectral sequence calculating differential graded cohomology
 with coefficients, see the lemma in 1.1.5, the $E_2$ term is the
 Lie algebra cohomology of $Hol(U)$ with coefficients $N(U)$ and the
 sequence collapses.$\square$

{\bf 2.2.7} Note that Kawazumi calculated the cohomology of $Hol(X)$
with coefficients in $n$-densities for an open Riemann surface
$X$. Taking into account this result (equation (9.7) p.701
\cite{Kaw}), we have completely solved the problem of the differential
graded cohomology of $\mathfrak{g}(X)$ with coefficients in
(differential graded) tensor densities for open Riemann surfaces.

\subsection{The cosimplicial version}

{\bf 2.3.1} Let us think of the tangent sheaf $Hol$  as a sheaf of Lie algebras
constituing an object in the derived category ${\cal D}^b(X)$ of
the category of bounded complexes of sheaves on the complex manifold
$X$. The objects $Hol$ and $\mathfrak{g}$ are isomorphic in ${\cal
D}^b(X)$. The Lie algebra structure on $Hol$ corresponds to the fact
that there is a cohomological resolution which is a sheaf of
differential graded Lie algebras.  
  
According to \cite{HinSch}, for any sheaf of Lie algebras
$\mathfrak{h}$ there is another sheaf of differential
graded Lie algebras constituing a resolution of $\mathfrak{h}$. It is the
sheaf of cosimplicial Lie algebras given by taking $\mathfrak{h}$ on
the Cech complex associated to a covering ${\cal U}$ by Stein open
sets, suitably normalised by the Thom-Sullivan functor, see \cite{HinSch}. 

{\bf 2.3.2} There is also a notion of cohomology for a cosimplicial
Lie algebra: the cohomology of the cosimplicial Lie algebra
$\check{\cal C}({\cal U},Hol)$ for some covering by Stein open sets
${\cal U}$ is the cohomology of the realization of the simplicial
cochain complex obtained from applying the continuous
Chevalley-Eilenberg complex as a functor $C^*_{cont}$ to the
cosimplicial Lie algebra. We denote cosimplicial cohomology by $H_{cos}^*$. 

{\bf 2.3.3} As explained in the introduction, the general idea is that
this cannot give anything new.

To show this, one constructs a morphism of simplicial cochain complexes

\begin{displaymath}
\tilde{f} : C^*_{dg}(\mathfrak{g}(N_{*}))\to C^*_{cont}(Hol(N_{*}))
\end{displaymath}

induced by the inclusion $f : Hol(N_{M,q})\hookrightarrow
\mathfrak{g}(N_{M,q})$ simply by applying the functor $C^*_{dg}$ to
the inclusion. $N_{*}$ denotes the thickened nerve of the covering
${\cal U}$, i.e. the simplicial complex manifold associated to the
covering ${\cal U}$. By lemma 5.9 in \cite{BotSeg}, the morphism $\tilde{f}$
induces a cohomology equivalence between the realizations of the two simplicial
cochain complexes (the conditions of the lemma are fullfilled because
of the isomorphism of the cohomologies on a Stein open set of
the covering and the K\"unneth theorem). As in prop. 6.2 in
\cite{BotSeg} using partitions of unity, one shows that the cohomology
of the realization of the simplicial cochain complex on the left hand
side gives the differential graded cohomology of $\Ga(X,\mathfrak{g})$.

{\bf 2.3.4} This gives the following

\begin{theo}
On a complex manifold $X$ of dimension $n$, we have

\begin{displaymath}
^1H^*_{dg}(\Ga(X,\mathfrak{g}))\cong H^*_{cos}(\check{C}({\cal U},Hol))
\end{displaymath}

for any covering of $M$ by Stein open sets ${\cal U}$.
\end{theo}

{\bf 2.3.5} Observe that we proceeded in the same order of taking
cohomology with respect to differentials in the spirit of remark 1.2.3. 

\section{Calculating the cohomology}

{\bf 3.1.1} I. M. Gelfand and D. B. Fuks calculated the cohomology of
the Lie algebra of formal vector fields in $n$ variables $W_n$ (in our
setting always with complex coefficients). They showed an isomorphism
of the Hochschild-Serre spectral sequence for the subalgebra $gl(n)$
with the Leray spectral of the restriction to the $2n$ squeleton of
the universal $U(n)$ principal bundle. 

Let us note $\pi: V(\infty,n)\to G(\infty,n)$ the universal principal
$U(n)$-bundle and $X(n)$  an open neighbourhood (because the inverse
image of the union of the cells is not a manifold) of the inverse
image under $\pi$ of the $2n$-squeleton of the Grassmannian
$G(\infty,n)$.

Their theorem reads

\begin{theo}[Gelfand-Fuks, cf \cite{Fuk}]
There is a manifold $X(n)$ such that

\begin{displaymath}
H^*_{cont}(W_n) \cong H^*_{sing}(X(n)).
\end{displaymath}
\end{theo}

R. Bott and G. Segal showed that for $R^n$ or more generally a
starshaped open set $U$ of an $n$-dimensional manifold $M$, the Lie algebra of
$C^{\infty}$-vector fields $Vect(U)$ has the same cohomology as $W_n$.

The same is true for the Lie algebra of holomorphic vector fields on a
disk of radius $R$ in $\C^n$: The map sending a holomorphic field to
its Taylor series is continuous (E. Borel's lemma, see \cite{Tre}
p.190), open (trivial !), injective (trivial !) and of dense image
(the series of convergence radius $R$ are dense in the formal
series). So they have the same continuous cohomology, cf \cite{Wag}.

{\bf 3.1.2} N. Kawazumi calcula\-ted what seemed to be the only interesting
Gel\-fand\--Fuks cohomology related to Lie algebras of holomorphic vector
fields on Riemann surfaces, i.e. the cohomology on open Riemann
surfaces:

\begin{theo}[Kawazumi, \cite{Kaw}]
Let $X$ an open Riemann surface. Then

\begin{displaymath}
H^*_{cont}(Hol(X)) = H^*(Map(X,S^3))
\end{displaymath}
\end{theo}

He used the method of Bott-Segal \cite{BotSeg} to prove this result,
i.e. he constructed a global fundamental map from the cochain complex
of the Lie algebra to the complex of differential forms with values in
$C^*_{cont}(Hol(\C))$. This map, denoted by $\hat{f}_{\sigma}$,  is
constructed with the help of a global non-vanishing vector field
$\partial$  existing on open Riemann surfaces:

\begin{eqnarray*}
\hat{f}_{\sigma} : C^*_{cont}(Hol(U_{\sigma}))\to
\Om^*(U^{\sigma};C^*_{cont}(Hol(\C)))\\
c\mapsto (\partial^{-1})\otimes (f_{\sigma,p})_* i_{\partial}c + (f_{\sigma,p})_*(c)
\end{eqnarray*}

Here, for a subset $\sigma = \{\alpha_0,\ldots,\alpha_q\}$ of the
index set of a covering, $U^{\sigma} = \bigcup_iU_{\alpha_i}$ and
$U_{\sigma} = \bigcap_iU_{\alpha_i}$. $(f_{\sigma,p})_*$ is the map
induced from a complex immersion of the open set into $\C$ and $i$ is
the insertion operator.

It is rather straight forward to generalize this map to the
$n$-dimensional case: $\hat{f}_{\sigma}$ relies on a vector valued
differential form $\om$ which is complicated in the case of Bott and
Segal, but here it is just $\om = \partial^{-1}\otimes \partial$, the
identity on $Hol(X)$. In the $n$-dimensional case, we take $\om =
\sum_{i=1}^n\partial^{-1}_i\otimes \partial_i$. These $\partial_i$ -
trivializing the tangent bundle - can be chosen such that they are
the images of $\frac{\partial}{\partial z_i}$ for a specially chosen
parametrization sending a contractible open set into $\C$, cf lemma
6.4 of \cite{Kaw}.   

{\bf 3.1.3} In general, there is no such vector field $\partial$, so there one
should adapt the fundamental map of Bott-Segal to this holomorphic
setting. For this, it is enough to notice that $X(n)$ is homotopically
equivalent to a complex manifold carrying a
$Gl(n,\C)$-action. For example, $X(1)$ is $S^3$ which is
homotopically equivalent to $C^2\setminus \{0\}$. So, replacing from
the real case the principal $U(n)$-bundle (associated to the tangent
bundle) by the principal $Gl(n,\C)$-bundle (associated to the complex
tangent bundle), one has a family of immersions $P$, cf \cite{BotSeg}
\S 4 and p. 295, which is parametrized by a complex manifold
($Gl(n,\C)$) and consists of complex immersions. This implies that
that the fundamental map, constructed from this family as in
\cite{BotSeg} \S 4, goes from (cochains on) holomorphic fields to
(holomorphic differential forms with values in cochains on)
holomorphic fields. 

{\bf 3.1.4} Secondly, Kawazumi uses the fact that the open Riemann surface is a
Stein manifold to pass from the cosimplicial cohomology to the
cohomology of the Lie algebra of global holomorphic fields. His method
works perfectly for $n$-dimensional Stein manifolds.

So there are two immediate corollaries:

\begin{cor}
Let $X$ be an $n$ dimensional complex Stein manifold with trivial
tangent bundle. Then we have

\begin{displaymath}
H^*_{cont}(Hol(X)) \cong H^*_{sing}(Map(X,X(n))).
\end{displaymath}
\end{cor}

If one drops the ``Stein'' hypothesis, it is perhaps not possible to
globalize the result, but one can stay with the cosimplicial
cohomology:

\begin{cor}
Let $X$ be an $n$ dimensional complex manifold with trivial
tangent bundle and ${\cal U}$ a covering of $X$ by Stein open
sets. Then we have 

\begin{displaymath}
H^*_{cos}(\check{C}({\cal U},Hol)) \cong H^*_{sing}(Map(X,X(n))).
\end{displaymath}
\end{cor}

From 3.1.3 follows on the other hand:

\begin{theo}
Let $X$ be an $n$-dimensional complex manifold. Then we have:

\begin{displaymath}
H^*_{cos}(\check{C}({\cal U},Hol)) \cong H^*_{sing}(\Ga(E_n)).
\end{displaymath}

Here, $E_n$ is the bundle with typical fiber homotopically equivalent
to $X(n)$ associated to the principal $Gl(N,\C)$-bundle on $X$ (gotten
from the complex tangent bundle of $X$).
\end{theo}
 
{\bf 3.1.5} For $\Ga(\Sigma,\mathfrak{g})$ in the case of a compact
Riemann surface $\Sigma$, we have Feigin's theorem (note that many
theorems in this article could be named ``Feigin's theorem''):

\begin{theo}
\begin{displaymath}
^1H^*_{dg}(\Ga(\Sigma,\mathfrak{g}))\cong H^*(Map(\Sigma,S^3))
\end{displaymath}
\end{theo}

{\it Proof:}

In our setting, this theorem follows from the above considerations
because the $(C^2\setminus\{0\})$-bundle (or the $S^3$-bundle) is
trivial:

The given $S^1$-representation in $SO(4)$ may be lifted to
$Spin(4)$ and this representation is used to view the bundle as associated
to a principal $Spin(4)$-bundle which is trivial
because of the existence of a section by obstruction theory combined with
dimension arguments.$\square$

Let us remark that one can calculate $H^*(Map(\Sigma,S^3))$ by
standard methods, and the result is given in Feigin's article. In
particular, $H^1(Map(\Sigma,S^3))$ is 1-dimensional, and fixing a generator
means fixing the central charge $c$ of a Virasoro type cocycle, cf \cite{Fei}. 
 
\section{Applications in conformal field theory}

{\bf 4.1.1} Feigin's article \cite{Fei} treats the applications in
conformal field theory. We will summarize them briefly, see \cite{Fei}
and \cite{BeiFeiMaz} for more informations. As complex
manifolds $X$, we take here compact Riemann surfaces $\Sigma$ of genus
$g\geq 2$. 

As we deal with homology in this section, we replace the sheaf of
holomorphic vector fields $Hol$ by the sheaf of algebraic vector
fields $Lie$. In view of the stated difficulties in globalizing these
vector fields, we take the cosimplicial version, cf \S 2.3.

{\bf 4.1.2} Let $p\in\Sigma$ a point. Following Feigin, let us choose
the covering of $\Sigma$ by a formal disk $U_2$ around $p$ (in order
to be able to take algebraic fields on it) and the Zariski open set $U_1 =
\Sigma\setminus\{p\}$. This means that $Lie(U_2)$ is the Lie algebra
of formal jets of vector fields at $p$, completed by the ideal defined
by $p$. A similar remark applies to $Lie(U_1\cap U_2)$. So,
$Lie(U_1)$, $Lie(U_2)$ and $Lie(U_1\cap U_2)$ form a cosimplicial Lie
algebra.

{\bf 4.1.3} As the choice of a generator for $^1H^1(KS(\Sigma))$ fixes
the central charge $c$ of a Virasoro type cocycle, cf {\bf 3.1.5}, it
fixes a cosimplicial Lie algebra associated to $Lie(U_1)$,
$Lie(U_2)\oplus c\C$ and $Vir(U_1\cap U_2)$ in the same way as
before. We have still inclusions of $Lie(U_1)$ and $Lie(U_2)\oplus
c\C$ into $Vir(U_1\cap U_2)$, because the cocycle is 0 on these
subspaces by the residue theorem. The cosimplicial Lie algebra is
denoted by $Lie_0(\Sigma)$. 

It has a representation (in the sense of representation of a diagram,
cf \cite{GerSch}) noted $\square_c$, where we associate to
$Lie(U_2)\oplus c\C$, $Vir(U_1\cap U_2)$ and $Lie(U_1)$ respectively
$1_c$ (a 1-dimensional space, $Lie(U_2)$ acting trivially, $c\C$
acting by multiplication by $c$), its induced module (a Verma module
noted $M_c(p)$) and its
restriction to $Lie(U_1)$. 

{\bf 4.1.4} There is a similar cosimplicial Lie algebra
$Lie_{\triangle}(\Sigma)$ associated to the covering by all Zariski
open sets of $\Sigma$. Such a set is given by finite number of points
$\{p_1,\ldots,p_n\}$. $Lie_{\triangle}(\Sigma)$ has a similar
representation: doing the above construction yields a representation
space for every $\Sigma\setminus\{p_1\}$. For Lie algebras associated
to sets with more than 1 point, we take the tensor product
representation of the Verma modules. Actually, all these modules are
linked by induction arrows. This gives a representation of
$Lie_{\triangle}(\Sigma)$ still noted $\square_c$. 

One should view $Lie_0(\Sigma)$ and its representation as a simple
model for $Lie_{\triangle}(\Sigma)$ and the above representation.

{\bf 4.1.5} Feigin calculates the (cosimplicial) homology of
$Lie_0(\Sigma)$ and $Lie_{\triangle}(\Sigma)$ with values in the above
representations. The result is (for simplicity only for $Lie_0(\Sigma)$)

\begin{theo}
\begin{displaymath}
H_i(Lie_0(\Sigma),\square_c) =\left\{ \begin{array}{c} 
M_c(p)\,\,/\,\,Lie(U_1)M_c(p)\,\,\,{\rm if}\,\,\,i=0 \\ 0\,\,\, {\rm
otherwise} \end{array}\right.
\end{displaymath}
\end{theo}

The point is that the space of coinvariants on the right hand side
which defines the so-called modular functor is usually associated to
locally defined objects, as for example the local Virasoro algebra
$Vir(U_1\cap U_2)$. Feigin obtaines here a homological description in terms of
globally defined objects.

A second point is that the space of coinvariants is in fact the continuous
dual of the completion of the local ring of the moduli space (of
compact Riemann surfaces of genus $g\geq 2$) at the point $\Sigma$,
provided that $\Sigma$ is a smooth point. This gives an important link between
Lie algebra homology and the geometry of the moduli space, cf \S 5 . 
    
{\bf 4.1.6} The modular functor for what is called a minimal field
theory relies on a special choice of the central charge $c$, dictated
from Virasoro representation theory, see for example \cite
{DiMaSe}. Furthermore, instead of Verma modules one deals with their
irreducible quotients. Feigin shows that the above setting can be
adapted to this situation. 

The modular functor associates to $\Sigma$ a finite dimensional vector
space; this fact relies in our context on the theorem, cf lemma
4.1.1 p. 16 \cite{FeiMal}, stating that coinvariants in a
representation with 0 singular support are finite dimensional.  

\section{Applications in deformation theory}

\subsection{Deformations of complex manifolds}

In this section, we give links from the cohomology calculations in the
first part to the deformation theory of complex manifolds, still
relying strongly on the ideas of \cite{Fei} and here also \cite{HinSch}.

It will concern particularly the differential graded homology of
$\Ga(X,\mathfrak{g})$ for a complex manifold $X$. The most part of
this section is more generally true for smooth proper schemes,
see \cite{HinSch}.

{\bf 5.1.1} The most basic idea in this context is the following, taken from
\cite{HinSch}: 

``The completion of a local ring of a moduli space at a given point $X$
is isomorphic to the dual of the $0^{\rm th}$ homology group of the
Lie algebra of infinitesimal automorphisms of $X$.''

Let me underline once more that this links Lie algebra homology and
the geometry of the moduli space in a formal neighbourhood of a point.

{\bf 5.1.2} As Feigin remarked, we have for Riemann surfaces an
incarnation of this principle: 

\begin{theo}
Let $\Sigma$ be a compact Riemann surface of genus $g\geq 2$. Then

\begin{displaymath}
^2H_{0,dg}(\Ga(\Sigma,\mathfrak{g})) = S^*(T_{\Sigma}{\cal M}(g,0))
\end{displaymath} 

and the other homology spaces are 0.
\end{theo}

{\it Remark:} Note that we have here $^2H_{0,dg}$; this reminds you of
the way we defined the differential graded homology of a sheaf of
differential graded Lie algebras, see 1.2.3.

{\it Proof:}

It is the result from the Kodaira-Spencer deformation theory for
Riemann surfaces $\Sigma$ that we have 

\begin{displaymath}
H^1(\Sigma,Hol) = T_{\Sigma}{\cal M}(g,0).
\end{displaymath}

Also, $H^0(\Sigma,Hol) = 0$. So the theorem follows directly from the
lemma in 1.1.6, because the graded Lie algebra homology of an abelian
Lie algebra in degree 1 is just the symmetric algebra on it.$\square$

Taking continuous duals in the theorem, we get the principle stated in
6.1.1 viewing $S^*(T_{\Sigma}{\cal M}(g,0))^*$ as the completion of the
local ring which is possible if the point $\Sigma$ is smooth in ${\cal
M}(g,0)$. 

{\bf 5.1.3} The theorem of 6.1.2 is still true for higher dimensional
complex manifolds $X$ as long as

\begin{equation}  \label{***}
H^1(X,Hol) =  T_{\Sigma}{\cal M}(g,0)
\end{equation}

and zero otherwise. So there are two problems, well known in deformation
theory following Kodaira and Spencer: the problem whether the number
of moduli is well-defined and the problem if equation \ref{***}
holds. For compact complex manifolds  $M$ this is answered by a
theorem of Kodaira, see \cite{Kod} p. 306 thm. 6.4: a sufficient
condition for the affirmative answer to the two questions is that

\begin{displaymath} 
H^0(M,Hol) = H^2(M,Hol) = 0.
\end{displaymath}

So in the case of compact complex surfaces, we can conclude right
away that the theorem in 6.1.2 is still true. See \cite{Kod} for
examples of such complex surfaces. 

\subsection{Deformations of Lie algebras}

{\bf 5.2.1} It is well known that the Lie algebra cohomology 
with values in the adjoint representation $H^*(L,L)$ of a Lie algebra
$L$ answers questions about the deformations of $L$ as an algebraic
object. For example, $H^2(L,L)$ can be interpreted as the space of
equivalence classes of infinitesimal deformations of $L$, see
\cite{Fuk} p. 35.

So there arise natural questions of this type for the Lie algebra of
holomorphic vector fields $Hol(U)$ on a Stein manifold $U$ and in the
differential graded setting for the differential graded Lie algebra
$\Ga(U,\mathfrak{g})$.

{\bf 5.2.2} The formal case is well known:

\begin{theo}
\begin{displaymath}
H^*_{cont}(W_n,W_n) = 0.
\end{displaymath}
\end{theo}

This gives right away (as before by considering $Hol(D)$ for a disk
$D\subset\C^n$ as a dense subalgebra of $W_n$ and by the principle
that a dense subalgebra has the same continuous cohomology)

\begin{cor}
\begin{displaymath}
H^*_{cont}(Hol(D),Hol(D)) = 0.
\end{displaymath}
\end{cor}

So this implies the rigidity of the Lie algebra of holomorphic vector
fields for disks. Observe that these disks are also rigid as
manifolds, i.e. $H^1(D,Hol) = 0$.

{\bf 5.2.3} Now by the theorem in 2.2.5, we also have differential
graded rigidity of $\Ga(D,\mathfrak{g})$

\begin{cor}
\begin{displaymath}
^1H^*_{dg}(\Ga(D,\mathfrak{g}),\Ga(D,\mathfrak{g})) = 0.
\end{displaymath}
\end{cor}

{\bf 5.2.4} On the other hand, for a compact Riemann surface $\Sigma$
of genus $g\geq 2$, we have by the lemma in 1.1.5 and by the exact
sequence which is implicit in the proof of the theorem in 5.1.2 (here,
we have the dg-cohomology procedure as in 5.1.2 !)

\begin{theo}
\begin{displaymath}
^2H^*_{dg}(\Ga(\Sigma,\mathfrak{g}),\Ga(\Sigma,\mathfrak{g})) =
S^*(T_{\Sigma}{\cal M}(g,0))^*\otimes T_{\Sigma}{\cal M}(g,0) .
\end{displaymath}
\end{theo}

Here, $S^*(T_{\Sigma}{\cal M}(g,0))^*$ is the continuous dual of the
nuclear Fr\'echet space given by the polynomials on $T_{\Sigma}{\cal
M}(g,0)$. So, it's the space of formal power series on $T_{\Sigma}{\cal
M}(g,0)^*$.
 
{\bf 5.2.5} Note that the space on the right hand side can be given a
bracket such that it is isomorphic to the Lie algebra of formal vector
fields on $T_{\Sigma}{\cal M}(0,g)$.

This could be interpreted as the relation between cohomology with
adjoint coefficients of $\mathfrak{g}$, i.e. differential graded
deformations of global sections of $\mathfrak{g}$, and deformations
of the underlying manifold. It fits into Feigin's philosophy that
the choice of the coefficients in the Lie algebra cohomology
determines the geometric object on the moduli space in a formal
neighbourhood of a point: trivial coefficients correspond to the
structure sheaf, adjoint coefficients correspond to vector fields,
adjoint coefficients in the universal envelopping algebra correspond to
differential operators.     

\section{Applications in foliation theory}

This section is inspired by the famous link between the cohomology of
Lie algebras and characteristic classes of foliations, see for example
\cite{Fuk} for an introduction. We won't go into
all details and we won't try to develop this theory in all its strength
in our case, alas, we will only consider the easiest case, i.e. the
case of characteristic classes of $g$-structures. In fact, we will
define a class of ``$g$''-structures such that the cohomology
calculations from the first part yield characteristic classes for
these structures. 

We won't pretend that this construction gives rise
to interesting new characteristic classes; in fact, in absence of an explicit
description of the cohomology classes, we have no explicit description
of the characteristic classes.

{\bf 6.1.1} A $g$-structure on a manifold $X$ is a $g$-valued
$C^{\infty}$-differential $1$-form $\omega$ satisfying the Maurer-Cartan
equation:

\begin{displaymath}
- [\om(\xi_1),\om(\xi_2)] = d\om(\xi_1,\xi_2).
\end{displaymath}

For a continuous cochain $c\in C^q_{cont}(g)$, there is a
characteristic class of the $g$-structure defined by $\omega$ simply given by the differential form

\begin{displaymath}
c(\underbrace{\omega,\ldots,\omega}_{\rm q-times}).
\end{displaymath}

{\bf 6.1.2} Define for a covering ${\cal U}$ by open sets a
``$Hol$-${\cal U}$-structure'' or short $Hol$-structure as follows: 

Let $X$ be a complex manifold and ${\cal U} = \{U_i\}_{i\in I}$ a covering of
$X$ by open sets such that $I$ is a countable directed index
set. Consider the sheaves $Hol$ and $Vect$ of holomorphic resp. $C^{\infty}$ vector fields on $X$. For an inclusion of open sets
$U\subset V$, we have restriction maps

\begin{displaymath}
\phi_{VU} : Hol(V)\to Hol(U)\,\,\,{\rm and}\,\,\,\psi_{VU} : Vect(V) \to Vect(U).
\end{displaymath}

A $Hol$-structure is now a $Hol(U_i)$-valued differential $1$-form
$\omega_{U_i}$ for every open set $U_i$ of ${\cal U}$ such that
it verifies the Maurer-Cartan equation and furthermore
for an inclusion $U\subset V$ we have 

\begin{displaymath}
\phi_{VU}(\omega_V(\xi)) =
 \omega_U(\psi_{VU}(\xi))
\end{displaymath}

for all $\xi\in Vect(V)$.

If $X$ is part of the covering and $Hol(X) = 0$, then the
$Hol$-structure is 0, so let us restrict to coverings not including
$X$.

{\bf 6.1.3} To have a link with better known structures in foliation
theory, let us restrict ourselves to coverings by contractible open
sets (such that intersections are contractible).

Let $X$ be of complex dimension $n$. By the obvious base change, we
can think of $W_{2n}$ as being generated by $\frac{\partial}{\partial
z_i}$ and $\frac{\partial}{\partial \bar{z}_i}$,
$i=1,\ldots,n$. Denote by $W_{2n}|_{hol}$ the Lie subalgebra of
$W_{2n}$ generated by the $\frac{\partial}{\partial z_i}$ for
$i=1,\ldots,n$.

  Given a $Hol$-structure associated such a covering, denoted by ${\cal
U}$, we have the  

\begin{lem}
The data $\{\omega_U\}_{U\in{\cal U}}$ is equivalent to a
$W_{2n}|_{hol}$-valued differential form $\omega$.
\end{lem}

So, for these coverings, $Hol$-structures are special cases of
$W_n$-structures, and their importance is clear, see for example
\cite{Fuk} Ch. 3.1.3 B $3^{\circ}$, p. 231.
  
{\bf 6.1.4} To such a structure (for which obviously only the
transverse structure of the foliation is relevant), we assign now
characteristic classes by considering not $H^*_{cont}(Hol(X))$ which
could be too small, but $\H^*(X,C^*_{cont}(Hol))$ or better
$H^*(|C^*_{cont}(\check{C}({\cal U},Hol))|)$ which co\"{\i}ncide by section 2. 

The $Hol$-structure is defined such that by inserting $p$-times
$\omega_{U_{i_0}\cap\ldots\cap U_{i_q}}$ into each $c\in
C^p_{cont}(\prod_{i_0<\ldots<i_q}Hol(U_{i_0}\cap\ldots\cap U_{i_q}))$, 
one constructs an element $\omega$ of the Cech-DeRham complex
associated to the covering ${\cal U}$ on $X$. By the standard theorem
saying that the Cech-DeRham complex calculates only DeRham cohomology,
this $\omega$ gives rise to a well-defined cohomology class $[\omega]$,
the characteristic class associated to the $Hol$-structure.

\end{document}


\title{Erratum}     
\author{Friedrich Wagemann \\
        Institut Girard Desargues -- ESA 5028 du CNRS\\
	Universit\'e Claude Bernard Lyon-I\\
        43, bd du 11 Novembre 1918 \\
	69622 Villeurbanne Cedex FRANCE\\
        tel.: +33.4.72.43.11.90\\
        fax:  +33.4.72.43.00.35\\
        e-mail: wagemann@desargues.univ-lyon1.fr}
\maketitle

Friedrich Wagemann: Differential Graded Cohomology and Lie Algebras of Holomorphic Vector Fields, {\it Commun. Math. Phys.} {\bf 208 } (1999), 521--540\\

The author apologizes for introducing a wrong new object in the above stated article. The sheaf ${\cal H}om_{\rm cont}(C_{*,dg}({\mathfrak{g}}),\C)$ of continuous homomorphisms of sheaves between the sheaf of differential graded chains $C_{*,dg}({\mathfrak{g}})$ of a sheaf of Lie algebras ${\mathfrak{g}}$ and the constant sheaf $\C$ is not a good object. It was intended to generalize continuous cochains of Lie algebras to a sheaf setting, but this is probably not possible for cochains with trivial coefficients. 

The space of global sections $Hom_{\rm cont}(C_{*,dg}({\mathfrak{g}}),\C)$ of ${\cal H}om_{\rm cont}(C_{*,dg}({\mathfrak{g}}),\C)$ is smaller than the space of continuous cochains $Hom_{\rm cont}(C_{*,dg}({\mathfrak{g}}(X)),\C)$, for example for the sheaf of differentiable vector fields $Vect$ on a finite dimensional compact manifold $X$. 

Indeed, examples of cochains (i.e. elements of $Hom_{\rm cont}(C_{*,dg}({\mathfrak{g}}(X)),\C)$) are evaluations $D_{x_0}(\xi_1,\ldots,\xi_r)=D(\xi_1,\ldots,\xi_r)(x_0)$ at a point $x_0\in X$ of some differential expression $D(\xi_1,\ldots,\xi_r)$ where $\xi_1,\ldots,\xi_r\in Vect(X)$. $D_{x_0}$ cannot be viewed as a morphism of sheaves, because $D_{x_0}$ is not restrictable to the open set $U^*=U\setminus\{x_0\}$, where $U$ is an open neighbourhood of $x_0$: 

\vspace{1cm}
\hspace{3cm}
\setlength{\unitlength}{1cm}
\begin{picture}(4,2)
\put(0,0){$\La^r(Vect(U^*))$} \put(3.8,0){$\C$}
\put(0,2){$\La^r(Vect(X))$} \put(3.8,1.8){$\C$}
\put(2.8,.1){\vector(1,0){.6}}\put(2.8,2){\vector(1,0){.6}}
\put(1,1.5){\vector(0,-1){.8}}\put(4,1.5){\vector(0,-1){.8}}
\put(1.1,1){${\rm res}_{\La^r(Vect)}$} \put(3.3,1){${\rm res}_{\C}$}
\put(3,1.6){$D_{x_0}$} \put(3,.3){$\phi$}
\end{picture}
\vspace{1cm}

But since ${\rm res}_{\C}={\rm id}_{\C}$, there does not exist a $\phi$ making the diagram commutative. Thus, the cochain $D_{x_0}$ is not an element of $Hom_{\rm cont}(C_{*,dg}({\mathfrak{g}}),\C)$.
 
A similar argument shows that cochains which are integrals over differential expressions are not elements of $Hom_{\rm cont}(C_{*,dg}({\mathfrak{g}}),\C)$. These two examples constitue the most important classes of cochains with values in the trivial module $\C$. 

Clearly, sheaf approaches are possible (and carried out) in the case of cochains with values in a module which is itself a non-constant sheaf, for example the sheaf of sections of a vector bundle.

In conclusion, one has to ban the object ${\cal H}om_{\rm cont}(C_{*,dg}({\mathfrak{g}}),\C)$ from the setting from the article (in particular, 1.1.8). The spectral sequence for continuous differential graded cohomology (lemma 1 to 4), the cosimplicial version (section 2.3, theorem 4) work still well, and the main result of the article (theorem 7) is unaffected.